\documentclass[aps,prl,twocolumn,showpacs]{revtex4}

\usepackage{graphicx}
\usepackage{epstopdf}
\usepackage{amssymb}
\usepackage{psfrag}
\usepackage{bm}
\usepackage{amsmath}
\usepackage{hyperref}
\usepackage{natbib}
\begin{document}

\title{Coherent Population Trapping with Controlled Interparticle Interactions}

\author{H. Schempp}
\author{G. G{\"u}nter}
\author{C.S. Hofmann}
\author{C. Giese}
\altaffiliation{Permanent address: Physikalisches Institut, Universit{\"a}t Freiburg, Hermann-Herder-Str. 3, 79104 Freiburg, Germany}
\author{S.D. Saliba}
\altaffiliation{Permanent address: Centre of Excellence for Coherent X-ray Science, School of Physics, University of Melbourne 3010, Australia}
\author{B.D. DePaola}
\altaffiliation{Permanent address: J.R. Macdonald Laboratory, Department of Physics, Kansas State University, Manhattan, Kansas 66506-2601, USA}
\author{T. Amthor}
\author{M. Weidem{\"u}ller}
\affiliation{Physikalisches Institut, Universit{\"a}t Heidelberg,
  Philosophenweg 12, 69120 Heidelberg, Germany}
\author{S. Sevin\c{c}li}
\author{T. Pohl}
\affiliation{Max Planck Institute for the Physics of Complex Systems, N\"othnitzer Strasse 38, 01187 Dresden, Germany}

\date{\today}

\begin{abstract}
We investigate Coherent Population Trapping in a strongly interacting ultracold Rydberg gas. Despite the strong van der Waals interactions and interparticle correlations, we observe the persistence of a resonance with subnatural linewidth at the single-particle resonance frequency 
as we tune the interaction strength. This narrow resonance cannot be understood within a meanfield description of the strong Rydberg--Rydberg interactions. Instead, a many-body density matrix approach, accounting for the dynamics of interparticle correlations, 
is shown to reproduce the observed spectral features. 
\end{abstract}

\pacs{42.50.Gy,42.50.Ct,32.80.Ee}
\maketitle

Coherent population trapping (CPT), i.e. the population of a quantum state decoupled from a resonant light field, serves as a paradigm for a quantum interference effect \cite{review_cpt}. First observed in 1976 \cite{gozzini1976}, CPT with its related phenomena electromagnetically induced transparency (EIT) \cite{harris1997,fleischhauer2005b} and stimulated Raman adiabatic passage (STIRAP) \cite{bergmann1998} has provided the basis for a large variety of effects and applications in many areas of physics, such as high-resolution spectroscopy, coherent control, metrology, quantum information and quantum gases. While CPT, EIT and STIRAP are generally described in terms of isolated single-atom interactions with coherent light fields, the situation becomes more involved when interactions between the particles need to be considered.

\begin{figure}
\includegraphics[scale=0.8]{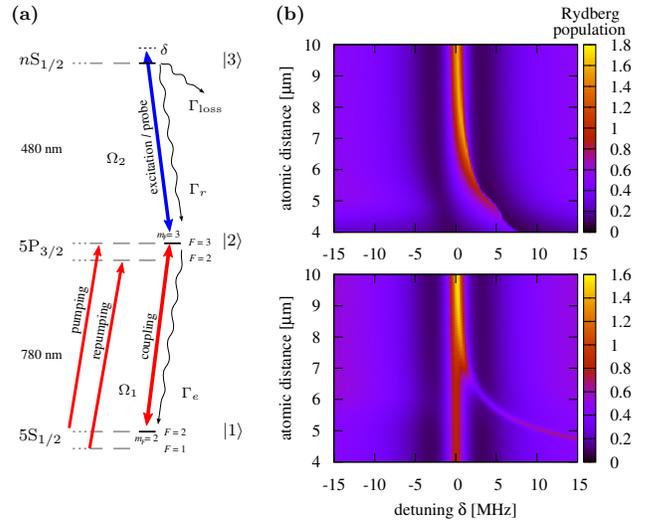}
 \caption{(a) Excitation scheme ($^{87}$Rb). $\Omega_1$ and $\Omega_2$ are the Rabi frequencies at 780 and 480\,nm, respectively, $\delta$ is the detuning of the upper transition; (b) calculated Rydberg state population, produced by the two-step sequence described in the text. The upper panel shows the result of a meanfield calculation, which predicts a strong shift and broadening of the resonance line. On the contrary, the exact result within a two-atom model (lower panel) yields an unshifted narrow interaction independent resonance and an additional resonance at $\delta=V(R)/h$ with $V(R)$ being the (repulsive) two-body interaction energy at an interparticle distance $R$.}
 \label{2atcalc}
\end{figure}

To gain initial insights into the effects of interactions on the quantum interference in CPT, consider two atoms with a three-level ladder structure with states $|1\rangle$, $|2\rangle$ and $|3\rangle$ as shown in Fig.~\ref{2atcalc}(a). The atoms are exposed to two resonant coherent light fields and interact only if both of them are in the highly excited atomic state $|3\rangle$. In the case of non-interacting atoms the population accumulates in the two-body product state of the single-particle dark state $|d\rangle$ which is a coherent superposition of $|1\rangle$ and $|3\rangle$ \cite{review_cpt}. This state is defined as the eigenstate of the total Hamiltonian with vanishing coupling to the coherent light field. When turning on the interparticle interaction this state is no longer a dark state as it is no longer an eigenstate of the total Hamiltonian. As pointed out in ~\cite{mmm08}, the two interacting atoms, nevertheless, possess two dark states $|d_\pm\rangle$. These states are dissipative due to the admixture of the intermediate, decaying state $|2\rangle$, but are significantly populated by optical pumping. While these states have dissipative character, they do not contain the state $|33\rangle$ and are, thus, immune to interactions.

In a first approach to a many-particle system one could apply a meanfield model by replacing many-body operators by products of their mean values thus  neglecting interparticle correlations. Fig.~\ref{2atcalc}(b) depicts the number of atoms in state $|3\rangle$ as a function of the upper laser detuning $\delta$ and the pair distance, and compares the meanfield result (upper panel) to a solution of the fully correlated two-atom Bloch equations (lower panel). The meanfield model predicts a shift and significant broadening of the resonance line, due to the energy shift and decoherence of state $|3\rangle$ induced by the meanfield interaction with the surrounding atoms. In contrast, the results of the two-atom model clearly indicate the population of the dark state as a non-shifted resonance without significant broadening. However, the states $|d_\pm\rangle$ are non-separable and can therefore not be properly accounted for in the meanfield model. The blue-shifted resonance line in the two-atom model can be assigned to the weakly populated eigenstate containing a contribution of $|33\rangle$ which is prone to interactions.

In this Letter we address the question to which extent a narrow unshifted CPT resonance persists in a multi-particle system with tunable interactions. More specifically, we investigate the three-level excitation of a gas of Rydberg atoms which are subjected to long-range van der Waals interactions in the blockade regime \cite{pillet2008,saffman2009,choi2007}. Rydberg atoms offer significant interparticle interactions over large distances which can be conveniently tuned by choosing the appropriate excited Rydberg state~\cite{singer2005b}. The effect of the interactions can further be controlled by changing the density of ground-state atoms in the gas. Interaction-induced dephasing \cite{raitzsch2009} and loss mechanisms \cite{weatherill2008} on light propagation in a Rydberg gas under EIT conditions have recently been studied. The work presented here provides a many-body approach which properly accounts for interparticle correlations induced by the long-range Rydberg interactions.

In our experimental scheme the states $|1\rangle$, $|2\rangle$ and $|3\rangle$ are represented by the atomic states 5S$_{1/2}(F=2,m_F=2)$, 5P$_{3/2}(F=3,m_F=3)$ and a Rydberg state \textit{n}S of $^{87}$Rb atoms which are coupled by resonant laser fields at 780\,nm and 480\,nm, respectively (see Fig.\,\ref{2atcalc}(a)). The ground state atoms are trapped in a magneto-optical trap at a maximum density of $\rho_0=6.6\times 10^9\,$cm$^{-3}$. This maximum density of atoms in state $|1\rangle$ can be reduced in a controlled way by optically pumping atoms into another hyperfine state that does not couple to the excitation laser fields as described in Ref.~\cite{reetz2008njp}. The Rydberg atoms are detected by field ionisation. Further details on the experimental setup can be found in Ref.~\cite{deiglmayr2006}.

We apply a double pulse excitation scheme where the first pulse resonantly excites up to 20\,\% of the ground state atoms to the Rydberg state. The resulting well-defined mixture of atoms in the ground state and Rydberg state is probed by scanning the blue laser frequency of the second pulse.  This initial partial excitation of the gas permits off-resonant excitation of strongly interacting atom pairs \cite{reinhard2008} during the second pulse, such that interaction effects are more pronounced. The first excitation pulse with a duration of $800\,$ns is realized by two circularly polarized counter-propagating laser beams, resonant with the respective transitions with peak Rabi frequencies of $\Omega_1=7.6\,$MHz and $\Omega_2=1.4 \,$MHz, respectively. The lower (red) excitation beam has a large beam radius ($\approx$1\,mm), while the upper (blue) beam is focussed to a waist of $\approx$37\,$\mu$m. The red Rabi frequency can thus be considered to be constant over the  narrow cylindrical excitation volume, while the blue Rabi frequency varies radially. After the first excitation pulse a second pair of laser pulses having the same beam geometry, but independently adjustable Rabi frequencies, probes the system with a pulse duration of $3\,\mu$s. While the lower laser transition is still resonant ($\Omega_1=2.7\,$MHz), the upper probe transition is scanned over the atomic resonance with a Rabi frequency of $\Omega_2=1.4$\,MHz. For the given parameters, on resonance the system is found close to the steady state after the probe pulse. A similar Rydberg excitation sequence has been employed in Ref.~\cite{reinhard2008} to probe energy shifts in a very dense sample with a detuned second excitation pulse. In contrast to our experiment, much shorter pulses were employed leading to a system far away from the steady-state.

\begin{figure}
 \includegraphics[scale=0.6]{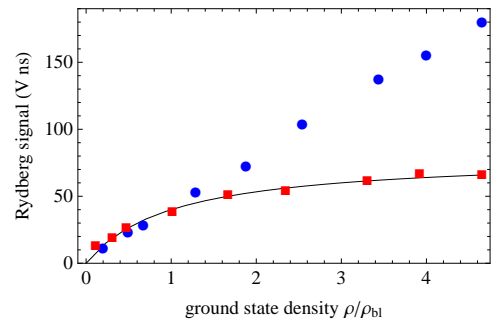}
 \caption{Excitation blockade of the Rydberg state 61S (filled squares) in comparison to the excitation of the 30S state (filled circles) after the first (resonant) excitation pulse. While the 30S state shows no saturation, the solid line in the 61S data shows a heuristic saturation function $\propto 1/(1+\rho_{\mathrm{bl}}/\rho)$ from which a blockade density $\rho_{\mathrm{bl}}\approx 1.3\times 10^9\,$cm$^{-3}$ is derived.}
\label{blockade}
\end{figure}

As a signature of interparticle interactions, the excitation blockade due to repulsive van der Waals interactions is presented in Fig.~\ref{blockade}. There is no excitation blockade observed for the 30S state reflecting the  $n^{11}$ dependence of the van der Waals interaction on the principal quantum number $n$~\cite{singer2005b}.

\begin{figure*}
 \includegraphics[scale=1.0]{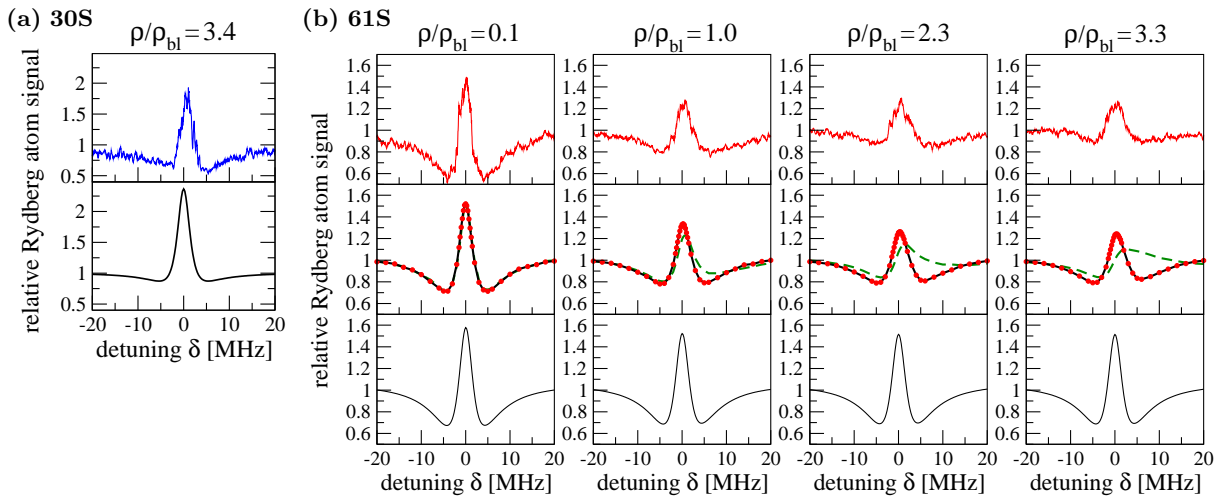}
 \caption{(a) Probe scan of the 30S Rydberg state (upper graph) and simulation using one-atom Optical Bloch Equations (lower graph), averaged over the Gaussian distribution of $\Omega_2$.
The 30S state is subject to stronger decay and redistribution than the 61S state, which is why the dip in the signal around the central peak is less pronounced.
 (b) Upper graphs: Similar scans for the 61S Rydberg state at different densities. All densities are given in relation to the density $\rho_{\mathrm{bl}}$ at which blockade effects become apparent. Middle graphs: Theoretical spectra, obtained from the density matrix expansion (red dots) and from the meanfield calculation (green dashed lines). Lower graphs: Theoretical spectra considering only interacting pairs.}
 \label{fig:spectra}
\end{figure*}

\begin{figure}
 \resizebox{0.9\columnwidth}{!}{\includegraphics{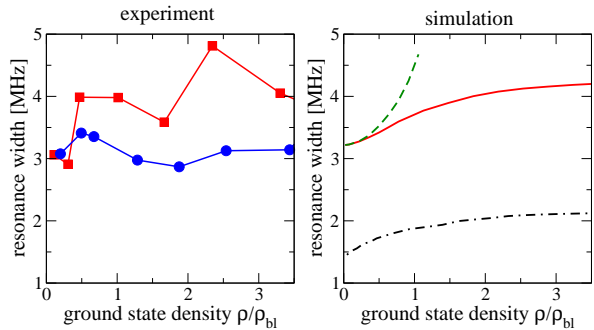}}
 \caption{The width of the narrow line in the measured 61S spectra (red squares) shows a weak density dependence and lies below the natural linewidth of the intermediate state at small densities. In comparison, the non-interacting 30S state does not show a significant change of the linewidth with density (blue circles). The many-body simulation (solid red curve) reproduces the slight increase of the linewidth for 61S whereas the meanfield model (dashed green line) predicts a much stronger density dependence. The black dash-dotted line shows the simulation results for vanishing laser linewidths. All values are FWHM determined from fitting the sum of two Gaussians of opposite sign.}
 \label{fig:linewidths}
\end{figure}

The corresponding CPT spectrum for the 30S state is shown in Fig.~\ref{fig:spectra}(a). The observed CPT resonance can be well described in terms of single-particle Optical Bloch Equations (OBEs) averaged over the Gaussian distribution of the Rabi frequency $\Omega_2$. The finite laser linewidth and redistribution of Rydberg states by blackbody radiation have been included as additional decay processes. Consistency of measurements with the prediction of the OBEs was confirmed for various pulse sequences and Rabi frequencies. Besides a scaling of the whole spectrum proportional to the density we do not find any density dependent features in the spectra. In the regime of strong Rydberg--Rydberg interactions (61S state with a van der Waals coefficient $C_6=-1.18 \times 10^{21}$ a.u. \cite{singer2005b}), as shown in Fig.~\ref{fig:spectra}(b) the CPT spectra still exhibit a pronounced peak with sub-natural linewidth at zero detuning at all densities. The width of the resonance slightly increases with density as depicted by the squares in Fig.~\ref{fig:linewidths}. The CPT resonance width of $\approx 3\,$MHz is well below the natural linewidth of the intermediate state $|2\rangle$ of 6.1\,MHz. As can be deduced from the comparison with the non-interacting 30S state (circles in Fig.~\ref{fig:linewidths}) the resonance width is mainly determined by the finite laser linewidths.

Our theoretical treatment of the excitation dynamics starts from the corresponding Heisenberg equations for the atomic population and transition operators $\hat{\sigma}_{\alpha\beta}^{(i)}=|\alpha_i\rangle\langle\beta_i|$ ($\alpha,\beta=1,2,3$), where $i=1,\ldots, N$ labels the $N$ atoms whose positions ${\bf r}_i$ are randomly sampled from the underlying density distribution. Taking expectation values one obtains equations of motion for the reduced single-atom, two-atom, etc. density matrices, $\rho_{\alpha\beta}^{(i)}=\langle\hat{\sigma}_{\alpha\beta}^{(i)}\rangle$, $\rho_{\alpha\beta,\alpha^{\prime}\beta^{\prime}}^{(i,j)}=\langle\hat{\sigma}_{\alpha\beta}^{(i)}\hat{\sigma}_{\alpha^{\prime}\beta^{\prime}}^{(j)}\rangle$, respectively. The additional interaction terms result in a hierarchy of coupled equations that ultimately requires knowledge of the $N$-atom density matrix for an exact solution and thus needs to be truncated in an appropriate way. The simplest possibility corresponds to the meanfield approximation, $\rho_{\alpha\beta,\alpha^{\prime}\beta^{\prime}}^{(i,j)}=\rho_{\alpha\beta}^{(i)}\rho_{\alpha^{\prime}\beta^{\prime}}^{(j)}+g_{\alpha\beta,\alpha^{\prime}\beta^{\prime}}^{(i,j)}\approx\rho_{\alpha\beta}^{(i)}\rho_{\alpha^{\prime}\beta^{\prime}}^{(j)}$, i.e. the neglect of direct two-particle correlations. Such meanfield treatments and slight modifications thereof \cite{tong2004,weimer2008c,chotia2008} have been successfully applied to model interaction-induced excitation suppression in cold Rydberg gases. While being appealingly simple, the resulting nonlinear, single-atom equations, however, imply a unphysical distance-independent level shift for any finite Rydberg excitation, which has more dramatic consequences for the shape of the excitation spectrum. As depicted as dashed lines in Fig.\,\ref{fig:spectra}(b) the meanfield treatment, incorporating the density profile as well as the spatial dependence of the Rabi frequencies of the experiment, fails once the excitation blockade sets in.

To account for two-atom entanglement within a many-body description, the density matrix approach is extended to second order, while approximately accounting for three-atom correlations. This is achieved by truncating the hierarchy via the two-atom cluster expansion $\rho_{\alpha\beta,\alpha^{\prime}\beta^{\prime},\alpha^{\prime\prime}\beta^{\prime\prime}}^{(k,i,j)}=\rho_{\alpha\beta}^{(k)}\rho_{\alpha^{\prime}\beta^{\prime},\alpha^{\prime\prime}\beta^{\prime\prime}}^{(i,j)}+\rho_{\alpha^{\prime}\beta^{\prime}}^{(i)}\rho_{\alpha\beta,\alpha^{\prime\prime}\beta^{\prime\prime}}^{(k,j)}+\rho_{\alpha^{\prime\prime}\beta^{\prime\prime}}^{(j)}\rho_{\alpha\beta,\alpha^{\prime}\beta^{\prime}}^{(k,i)}-2\rho_{\alpha\beta}^{(k)}\rho_{\alpha^{\prime}\beta^{\prime}}^{(i)}\rho_{\alpha^{\prime\prime}\beta^{\prime\prime}}^{(j)}+g_{\alpha\beta,\alpha^{\prime}\beta^{\prime},\alpha^{\prime\prime}\beta^{\prime\prime}}^{(k,i,j)}
$. Initially neglecting direct three-atom correlations (i.e. $g_{\alpha\beta,\alpha^{\prime}\beta^{\prime},\alpha^{\prime\prime}\beta^{\prime\prime}}^{(k,i,j)}$), one obtains a closed set of $37$ dynamical equations for each pair of atoms. The dynamics of the two-atom density matrices involves three-atom interaction terms of the type
\begin{eqnarray}\label{eq2}
\sum_{k\neq i,j}V_{jk}\rho_{\alpha\beta,\alpha^{\prime}\beta^{\prime},\alpha^{\prime\prime}\beta^{\prime\prime}}^{(k,i,j)}=\sum_{k\neq i,j}\left[V_{jk}\rho_{\alpha^{\prime}\beta^{\prime}}^{(i)}\rho_{\alpha\beta,\alpha^{\prime\prime}\beta^{\prime\prime}}^{(k,j)}\right.&&\nonumber\\
+\left.V_{jk}\rho_{\alpha\beta}^{(k)}g_{\alpha^{\prime}\beta^{\prime},\alpha^{\prime\prime}\beta^{\prime\prime}}^{(i,j)}+V_{jk}\rho_{\alpha^{\prime\prime}\beta^{\prime\prime}}^{(j)}g_{\alpha\beta,\alpha^{\prime}\beta^{\prime}}^{(k,i)}\right]\;.&&
\end{eqnarray}
The second and third terms vanish rapidly if one of the three atoms is farther apart than the blockade radius from the remaining pair, but diverge if all three atoms are simultaneously very close to each other. This unphysical small-distance behavior results from neglecting direct three-atom correlations, which would cancel the respective terms. To account for this fact in a simple way we disregard the second and third interaction term in eq.(\ref{eq2}). This procedure, corresponding to the so-called ladder approximation in kinetic theory \cite{kintheo}, properly accounts for mutual interactions of close and distant particles, and is approximate only if the mutual distances of more than two atoms are simultaneously on the order of the blockade radius.

The solid dots in the middle graphs of Fig.~\ref{fig:spectra}(b) show the results of this model
qualitatively reproducing the experimental data. The calculation reproduces the weak density dependence of the resonance width (solid line Fig.~\ref{fig:linewidths}) regardless of the strong excitation suppression observed for densities
$\rho>\rho_{\mathrm{bl}}$ (see Fig.~\ref{blockade}). Again this stands in pronounced contrast to the strong quadratic broadening obtained from the meanfield calculations. The simulations also reveal that the measured resonance widths are largely limited by the spectral width of the excitation lasers as seen from comparison with the dash-dotted line in Fig.~\ref{fig:linewidths} giving the genuine CPT resonance width.

The persistence of the narrow resonance may be qualitatively understood from the two-atom dark state~\cite{mmm08} discussed above. In order to draw a connection to this simplified, yet more intuitive, picture and to elucidate the role of many-particle effects, we also consider the excitation dynamics for pairs of interacting atoms. Assuming that the interaction shift of each atom is solely determined by its nearest neighbor, one may calculate the pair-spectrum for variable distances, averaged over the nearest neighbor distribution of the random gas. As shown in Fig.~\ref{fig:spectra} the pair model provides the basic mechanism for the interaction-resistent narrow resonance. As may be anticipated from Fig.~\ref{2atcalc}, the predicted linewidth becomes, however, density-independent for $\rho>\rho_{\rm bl}$, in contrast to the experimental observation. This is due to the oversimplified neglect of fluctuating, simultaneous interactions between several atoms, and more importantly interactions between blocked pairs, or larger numbers, of atoms. The improved agreement obtained using our reduced density matrix approach, that accounts for these processes, thus, highlights the importance of many-body effects.

In conclusion, we have presented an experimental scheme to investigate the role of interactions in a CPT scheme. As suggested by the occurrence of an entangled dark state in a two-body model, the interaction-induced correlations between the atoms cannot be accounted for by a meanfield model, and require a more sophisticated many-body theory. The theoretical framework developed in this work reproduces the observed density-dependent features of CPT resonances and complements current numerical methods that are limited to small samples and very small numbers of Rydberg excitations \cite{weimer2008c,younge2009,pohl2010}. The concepts of CPT and EIT in systems exhibiting Rydberg interactions are currently attracting much interest in the context of quantum information, quantum simulation and sensitive probing of electric fields \cite{mmm08,mohapatra2008b,muller2009,weimer2009}. Therefore the understanding of interaction effects is of crucial importance. The presented theory, supported by the reported experiments, may serve as a valuable basis for future studies of light propagation and EIT in interacting media, as well as addressing questions concerning optical nonlinearities and photon correlations induced by interparticle interactions.

\textit{Note added}.-- While this article was prepared for submission we have become aware of complementary work \cite{pritchard2009}, where the effect of interactions on EIT in a Rydberg gas is mapped onto the light field, leading to a cooperative optical non-linearity.

The authors acknowledge support by the Deutsche Forschungsgemeinschaft (grant no. WE2661/10-1) and the Heidelberg Center for Quantum Dynamics. We thank A. Ekers for fruitful discussions.

\end{document}